\documentclass[11pt]{article}
   \usepackage{amsmath,amssymb}
   \usepackage[round]{natbib}
   \usepackage[spanish,british]{babel}
   \usepackage{fancyhdr}
   \usepackage{geometry}
   \usepackage{graphicx}
\makeatletter

\def\@makechapterhead#1{\thispagestyle{empty}%
  \vspace*{50\p@}%
  {\parindent \z@ \raggedright \normalfont
    \ifnum \c@secnumdepth >\m@ne
      \if@mainmatter
        \huge\bfseries\centerline{ \@chapapp\space \thechapter}
        \par\nobreak
        \vskip 20\p@
      \fi
    \fi
    \interlinepenalty\@M
    \Huge \bfseries \centerline{#1}\par\nobreak
    \vskip 40\p@
  }}

\if@titlepage
  \newenvironment{resumen}{%
      \titlepage
      \null\vfil
      \@beginparpenalty\@lowpenalty
      \begin{center}%
        \bfseries \resumenname
        \@endparpenalty\@M
      \end{center}}%
     {\par\vfil\null\endtitlepage}
\else
  \newenvironment{resumen}{%
      \if@twocolumn
        \section*{\resumenname}%
      \else
        \small
        \begin{center}%
          {\bfseries \resumenname\vspace{-.5em}\vspace{\z@}}%
        \end{center}%
        \quotation
      \fi}
      {\if@twocolumn\else\endquotation\fi}
\fi
\newcommand\resumenname{Resumen}

\makeatother

\fancypagestyle{plain}{
			\fancyhead{}
			
}
\newcounter{oldsection}
\makeatletter
\@addtoreset{equation}{section}
\@addtoreset{table}{section}
\makeatother

\title{Hydrodynamical interaction between a shock wave and a cloud. One
       dimensional approach.}
\author{S. Mendoza \\
        Cavendish Laboratory, Madingley Rd., Cambridge CB3 OHE, U.K.\\
        (sergio@mrao.cam.ac.uk)}

\begin{document}

\maketitle

%%%%%%%%%%%%%%%%%%%%%%%%%%%%%%%%%%%%%%%%%%%%%%%%%%%%%%%%%%%%%%%%%%%
%%%%%%%%%%%%%%%%%%%%%%%%%A B S T R A C T%%%%%%%%%%%%%%%%%%%%%%%%%%%
%%%%%%%%%%%%%%%%%%%%%%%%%%%%%%%%%%%%%%%%%%%%%%%%%%%%%%%%%%%%%%%%%%%

\selectlanguage{spanish}
   \begin{resumen}
        La colisi\'on de una onda de choque plano paralela con una nube
      plano paralela de densidad uniforme es analizada para el caso
      en el que campos magn\'eticos y perdidas por radiaci\'on no son
      consideradas.  Se discuten soluciones anal\'{\i}ticas generales
      para el caso en el que la densidad de la nube es mucho mayor que
      la del gas que le rodea.  Este problema generaliza uno de los
      problemas cl\'asicos en el estudio de la din\'amica de gases:
      la colisi\'on entre una onda de choque y una pared s\'olida.
   \end{resumen}
\selectlanguage{british}

\begin{abstract}
   The collision of a plane parallel shock wave with a plane parallel
cloud of uniform density is analysed for the case in which magnetic fields
and radiative losses are not considered.  General analytic solutions are
discussed for the case in which the density of the cloud is greater than
that of the surrounding environment.  This problem generalises one of
the classical problems in gas dynamics: the collision between a shock
wave and a solid wall.
\end{abstract}


%%%%%%%%%%%%%%%%%%%%%%%%%%%%%%%%%%%%%%%%%%%%%%%%%%%%%%%%%%%%%%%%%%%%
%%%%%%%%%%%%%%%%%%%%%%%%%%%%%%%%%%%%%%%%%%%%%%%%%%%%%%%%%%%%%%%%%%%%
%%%%%%%%%%%%%%%%%%%%%%%%%%%%%%%%%%%%%%%%%%%%%%%%%%%%%%%%%%%%%%%%%%%

\section{Introduction}

   The problem of the collision of a shock wave with a cloud has been
intensely investigated in the past by several authors (see for example
\citealt{klein94} and references therein).  The simplest assumption to
make is to consider a cloud for which gravitational effects are not
considered, magnetic fields are non-important and radiative losses
are negligible.  The fact that gravity is not taken into account,
makes it possible to consider the density of the cloud as uniform.
The complete 3D hydrodynamical problem is extremely complicated, even
under the simplifications mentioned above. However, numerical simulations
have been done for this case which ultimately give rise to instabilities
causing a complete disruption of the cloud \citep{klein94}.

   This article describes how the solution of the one dimensional
problem can be obtained. It has been argued in the past that at least for
the case in which the density contrast is high, i.e. the ratio of the
cloud's density to that of the external environment is high, the problem
has to be very similar to the one found in the problem of a collision
of a plane parallel shock with with a solid wall (\citealt{spitzer82},
\citealt{mckee88}).

   Many Astrophysical phenomena give rise to collisions between a
shock wave and a cloud.  For example, when a supernova explosion
occurs, the intense ejection of energy from the supernova into the
interstellar medium produces a spherical shock wave which expands into the
interstellar medium.  Several examples exist for which collisions of this
expanding shock have been observed to interact with clouds embedded in the
interstellar medium. This interaction is very important, since it seems to
induce, under not very well known circumstances, gravitational collapse
and star formation \citep{scientific}.  Another scenario is presented
by the expansion of jets around active galactic nuclei. A pair of jets
expand in opposite directions from the nuclei of the galaxy creating
a bow shock which interacts with the intergalactic medium.  It is the
interaction of this expanding bow shock with clouds or galaxies embedded
in cluster of galaxies that provides a mechanism in which shock--cloud
interactions take place.  It seems that this interaction is able to induce
star formation very efficiently\footnote{ See for example the Hubble Space
Telescope WWW site at http://oposite.stsci.edu/pubinfo/pr/1995/30.html}.

   Having all this considerations in mind, the present paper aims to give
a simple way of solving a particular case of the whole problem.  This
article provides an analytic description of the one dimensional problem
of a collision between a plane parallel shock with a plane parallel
``cloud'' bounded by two tangential discontinuities.  It is assumed
that the specific volume in the cloud is a quantity of the first order,
in other words  solutions are given for the case in which the density
of the cloud is much greater than that of the surrounding environment.

\section{General Description of the problem}

   Consider two plane parallel infinite tangential discontinuities.
The cloud, or internal region to the tangential discontinuities
has uniform pressure $p_c$ and density $\rho_c$.  The environment,
or external region to the cloud has also uniform values of pressure
$p_1$ and density $\rho_1$ respectively.  A plane parallel shock wave
is travelling in the positive $x$ direction and eventually will collide
with the left boundary of the cloud at time $t \! = \!  t_0 \! < \!  0$.
For simplicity we assume from now on that the density of the cloud is
greater than that of the environment.  By knowing the pressure $p_2$
and density $\rho_2$ behind the shock wave, it is possible to solve the
hydrodynamical problem thus defined.

   The problem of the collision of a shock wave and a tangential
discontinuity is well known \citep{daufm}.  Since at the instantaneous
time of collision the values of, say, the density in front and behind
the shock are $\rho_c$ and $\rho_2$ respectively, the standard jump
conditions for a shock no longer hold. A discontinuity in the initial
conditions (first initial discontinuity) occurs.

   When a discontinuity in the initial conditions occurs, the values
of the hydrodynamical quantities need not to have any relation at all
between them at the surface of discontinuity.  However, certain relations
need to be valid in the gas if stable surfaces of discontinuity are to be
created.  For instance, the Rankine-Hugoniot relations have to be valid in
a shock wave.  What happens is that this initial discontinuity splits into
several discontinuities, which can be of one of the three possible types:
shock wave, tangential discontinuity or weak discontinuity.  This newly
formed discontinuities move apart from each other with respect to the
plane of formation of the initial discontinuity.

   Very general arguments show that only one shock wave or a pair of
weak discontinuities bounding a rarefaction wave can move in opposite
directions with respect to the point in which the initial discontinuity
took place.  For, if two shock waves move in the same direction, the shock
at the front would have to move, relative to the gas behind it, with a
velocity less than that of sound.  However, the shock behind must move
with a velocity greater than that of sound with respect to the same gas.
In other words, the leading shock will be overtaken by the one behind.
For exactly the same reason a shock and a rarefaction wave can not move in
the same direction, and this is due to the fact that weak discontinuities
move at the velocity of sound relative to the gas they move through.
Finally, two rarefaction waves moving in the same direction can not
become separated, since the velocities of their boundaries with respect
to the gas they move through is the same.

  Boundary conditions demand that a tangential discontinuity must remain
in the point where the initial discontinuity took place.  This follows
from the fact that the discontinuities formed as a result of the initial
discontinuity  must be such that they are able to take the gas from a
given state at one side of the initial discontinuity to another state
in the opposite side.  The state of the gas in any one dimensional
problem in hydrodynamics is given by three parameters (say the pressure,
the density and the velocity of the gas).   A shock wave however, is
represented by only one parameter as it seen from the shock adiabatic
relation (Hugoniot adiabatic) for a polytropic gas:

\begin{equation}
   {\frac{V_b}{V_f}} = { \frac{(\gamma+1)p_f + (\gamma-1)p_b}
                       {(\gamma-1)p_f + (\gamma+1)p_b} }, 
\label{eq.2}
\end{equation}	

\noindent where $p$ and $V$ stand for pressure and specific volumes
respectively, $\gamma$ is the polytropic index of the gas and the
subscripts $f$ and $b$ label the flow in front of and behind the shock.
For a given thermodynamic state of the gas ( i.e.  for given $p_f$ and
$V_f$) the shock wave is determined completely since, for instance, $p_b$
would depend only on $V_b$ according to the shock adiabatic relation.
On the other hand, a rarefaction wave is also described by a single
parameter.  This is seen from the equations which describe the gas inside
a rarefaction wave which moves to the left with respect to gas at rest
beyond its right boundary \citep{daufm}:

\begin{gather}
  c_R = c_4 + {\frac{1}{2}} (\gamma_c-1) w_R,             \label{eq.4.a} \\
  \rho_R = \rho_4 \left\{ 1 + { \frac{1}{2}} \frac{(\gamma_c
     -1) w_R}{ c_4} \right\} ^ { 2 / (\gamma_c -1) },     \label{eq.4.b}\\
   p_R = p_4 \left\{ 1 + { \frac{1}{2}} \frac{(\gamma_c -1)
      w_R}{ c_4} \right\} ^ { 2\gamma_c / (\gamma_c -1)}, \label{eq.4.c}\\
   w_R = { -{ \frac{2}{\gamma_c +1} } \left(c_4 + {\frac{x}{t}}
       \right) }.                                         \label{eq.4}
\end{gather}

\noindent where $c_4$ and $c_R$ represent the sound speed behind and
inside the rarefaction wave respectively. The magnitude of the velocity of
the flow inside the rarefaction wave is $w_R$ in that system of reference.
The quantities $p_4$ and $p_R$ are the pressures behind and inside the
rarefaction wave respectively.  The corresponding values of the density
in the regions just mentioned are $\rho_4$ and $\rho_R$.

  With only two parameters at hand, it is not possible to give a
description of the thermodynamic state of the gas. It is the tangential
discontinuity, which remains in the place where the initial discontinuity
was produced, that accounts for the third parameter needed to describe
the state of the fluid.

  When a shock wave hits a tangential discontinuity, a rarefaction
wave can not be transmitted to the other side of the gas bounded by
the tangential discontinuity.  For, if there would be a transmitted
rarefaction wave to the other side of the tangential discontinuity,
the only possible way the boundary conditions could be satisfied is
if a rarefaction wave is reflected back to the gas.  In other words,
two rarefaction waves separate from each other in opposite directions
with respect to the tangential discontinuity that is left after the
interaction.  In order to show that this is not possible, consider a
shock wave travelling in the positive $x$ direction, which compresses
gas $1$ into gas $2$ and collides with a tangential discontinuity. After
the interaction two rarefaction waves separate from each other and
a tangential discontinuity remains between them.  In the system of
reference where the tangential discontinuity is at rest, the velocity
of gas $2$ is $v_2 \! = \! - \int_{p_3}^{p_2} \sqrt{ -{\rm d}p{\rm d}V
}$, where $p_3$ is the pressure of gas $3$ surrounding the tangential
discontinuity.  Accordingly, the velocity of gas $1$ in the
same system of reference is $v_1 \! = \! - \int_{p_3}^{p_1} \sqrt{ -{\rm
d}p{\rm d}V }$.  Since the product $-{\rm d}p{\rm d}V$ is a monotonically
increasing function of the pressure and $ 0 \le p_3 \le p_1 $ then:

\[
  -\int_0^{p_2}  \sqrt{ - {\rm d}P {\rm d}V } \le v_1 - v_2 \le \int_0^{p_1}
     \sqrt{ - {\rm d}P {\rm d}V } - \int_{p_1}^{p_2} \sqrt{ - {\rm d}P 
   {\rm d}V }. 
\]

\noindent The difference in velocities $v_1 - v_2$ has the same value in
any systems of reference and so, it follows that $ v_1 \! \le \!  v_2$,
in particular on a system of reference with the incident shock at
rest. However, for the incident shock to exist, it is necessary that
$v_1 \! > \!  v_2$, so two rarefaction waves can not be formed as a
result of the interaction.

  So far, it has been shown that after the collision between the shock
and the boundary of the cloud, a first initial discontinuity is formed.
This situation can not occur in nature in any manner and the shock splits
into a shock which penetrates the cloud and either one of a shock, or
a rarefaction wave (bounded by two weak discontinuities) is reflected
from the point of collision.  With respect to the point of formation of
the initial discontinuity, boundary conditions demand that a tangential
discontinuity must reside in the region separating the discontinuities
previously formed.

  In a shock wave, the velocities ($v$) in front and behind the shock are
related to one another by their difference:

\begin{equation}
   {v_f-v_b}=\sqrt{(p_b-p_f)(V_f-V_b)},
\label{eq.3}
\end{equation}

\noindent where the subscripts $f$ and $b$ label the flow of the gas
in front and behind the shock wave. 

\begin{figure}
   \begin{center}
       \includegraphics[height=5.4cm]{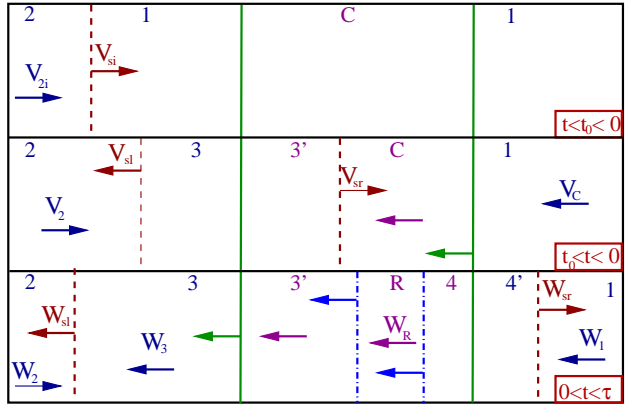}
   \end{center}
    \caption{ \footnotesize An incoming shock travelling to the right (top
             panel) hits a tangential discontinuity at time $t_0\!<\!0$.
             This produces two shocks moving in opposite directions with
             respect to the place of formation (middle panel).  When the
             transmitted shock into the cloud (region C) collides with
             its right boundary a reflected rarefaction wave (region
             R) bounded by two tangential discontinuities and a shock
             transmitted to the external medium (lower panel) are formed.
             Arrows represent direction of different boundaries, or the
             flow itself.  The numbers in the figure label different
             regions of the flow. Dashed lines represent shocks, dash-dot
             are weak discontinuities and continuous ones are tangential
             discontinuities.  The system of reference is chosen such
             that the tangential discontinuities which are left as a
             result of the collisions are always at rest.}
      \label{fig.1}
\end{figure}

  If after the first initial discontinuity two shock waves separate with
respect to the point of collision, then according to eq.(\ref{eq.3})
the velocities of their front flows are given by $ v_c \! = \! -\sqrt{
( p_3 - p_1 ) ( V_c - V_{3'} ) }$ and $v_2 \! = \! \sqrt{ ( p_3 - p_2 )
( V_2 - V_3 ) }$, where the regions $3$ and $3'$ bound the tangential
discontinuity which is at rest in this particular system of reference
(see top and middle panels of fig.(\ref{fig.1})).  Due to the fact that
$p_3 \ge p_2$ and because the difference $v_2 - v_c$ is a monotonically
increasing function of the pressure $p_3$, then:

\[
v_2 - v_c > ( p_2 -p_1 ) \surd  \left\{2V_c / \left[ \left( \gamma_c - 1 
   \right) p_1 + \left( \gamma_c + 1 \right) p_2 \right] \right\},
\]

\noindent according to the shock adiabatic relation.  Since $v_2 \! -
\! v_c$ is given by eq.(\ref{eq.3}), then:

\begin{equation}
  { \frac{V_1}{(\gamma-1)+(\gamma+1)p_2/p_1}} > {
  \frac{V_c}{(\gamma_c-1)+(\gamma_c+1)p_2/p_1} },
\label{eq.1}
\end{equation}

\noindent where $\gamma$ and $\gamma_c$ represent the polytropic indexes
of the environment and the cloud respectively.  $V_1$ and $V_c$ are
the specific volumes on the corresponding regions.  In other words,
a necessary and sufficient condition for having a reflected shock
from the boundary of the two media, under the assumption of initial
pressure equilibrium between the cloud and the environment, is given by
eq.(\ref{eq.1}).  Since for the problem in question $V_1\!>\!V_c$ and
the polytropic indexes are of the same order of magnitude, a reflected
shock is produced.

%%%%%%%%%%%%%%%%%%%%%%%%%%%%%%%%%%%%%%%%%%%%%%%%%%%%%%%%%%%%%%%%%%%%
%%%%%%%%%%%%%%%%%%%%%%%%%%%F I  G U R E 1%%%%%%%%%%%%%%%%%%%%%%%%%%%%
%%%%%%%%%%%%%%%%%%%%%%%%%%%%%%%%%%%%%%%%%%%%%%%%%%%%%%%%%%%%%%%%%%%%
%%%%%%%%%%%%%%%%%%%%%%%%%%%%%%%%%%%%%%%%%%%%%%%%%%%%%%%%%%%%%%%%%%%%
%%%%%%%%%%%%%%%%%%%%%%%%%%%F I G U R E 1%%%%%%%%%%%%%%%%%%%%%%%%%%%%
%%%%%%%%%%%%%%%%%%%%%%%%%%%%%%%%%%%%%%%%%%%%%%%%%%%%%%%%%%%%%%%%%%%%

   In the same form, at time $t\!=\!0$ when the transmitted shock reaches
the right tangential discontinuity located at $x \! = \! 0$, another
(second) initial discontinuity must occur.  In this case, we must invert
the inequality in eq.(\ref{eq.1}), change $\gamma$ by $\gamma_c$ and
$p_2$ by $p_3$, where $p_3$ is the pressure behind the shocks produced
by the first initial discontinuity.  Again, using the same argument for
the polytropic indexes, it follows that after this interaction a weak
discontinuity bounded by two rarefaction waves must be reflected from
the boundary between the two media.  As a result of the interaction,
once again, boundary conditions demand that a tangential discontinuity
remains between the newly formed discontinuities.

   This situation continues until the rarefaction wave and the left
tangential discontinuity of the cloud collide at time $t\!=\!\tau\!>\!0$.
At this point, two rarefaction waves separating from each other from
the point of formation will be produced once a stationary situation is
reached, and a tangential discontinuity will be separating the newly
formed discontinuities.  One can continue in a somewhat indefinite manner
with the solution but, for the sake of simplicity the calculations are
stopped at this point.  Fig.(\ref{fig.1}) shows a schematic description
of the solution described above in a system of reference such that the
tangential discontinuities which are left as a result of the different
interactions are at rest.  The numbers in the figure label different
regions in the flow.

\section{First initial discontinuity}

   According to fig.(\ref{fig.1}), after the first initial discontinuity
the absolute values of the velocities ($v$) of the flow are related by:

\begin{equation}
   v_2 + v_c = v_{2i}.
\label{eq.5}
\end{equation}

\noindent With the aid of eq.(\ref{eq.3}), the velocities of
eq(\ref{eq.5}) are given by:

\begin{gather}
   \lefteqn{ v_{2i}^2 = (p_2-p_1)(V_1-V_2), } \label{eq.6.a} \\
   \lefteqn{ v_c^2 = (p_3-p_1)(V_c-V_{3'}), } \label{eq.6.b} \\
   \lefteqn{ v_2^2 = (p_3-p_2)(V_2-V_{3}).  }  \label{eq.6}
\end{gather}

   Inserting eqs.(\ref{eq.6.a})-(\ref{eq.6}) into eq.(\ref{eq.5})
and substituting for the specific volumes from eq.(\ref{eq.2}), one
ends with a relation which relates the pressure $p_3$ as a function of
$p_2$, $p_1$ and the polytropic indexes in an algebraic linear form.
Straightforward manipulations show that the resulting equation does not
have an easy analytic solution, even for the particular cases in which
a strong or weak incident shock collides with the cloud.

   In order to find a set of analytic solutions, let us first describe
a particular solution to the problem.  If we consider a cloud with an
initial infinite density -a solid wall, then eq.(\ref{eq.5}) takes the
form $v_2 \!  =\!  v_{2i}$, and a ``zeroth order'' solution is found
\citep{daufm}:

\begin{equation}
   \frac{ p_{3_0} }{p_2} = { \frac{ (3\gamma-1)p_2 - (\gamma-1)p_1 }{
			          (\gamma-1)p_2 + (\gamma+1)p_1 }  },	
\label{eq.7}
\end{equation}

\noindent where $p_{3_0}$ is the value of the pressure behind the
reflected and transmitted shocks for the case in which the cloud has
specific volume $V_c \! = \! 0$.  For this particular case,
eq.(\ref{eq.7}) determines $p_{3_0}$ as a function of $p_1$ and $p_2$,
which are initial conditions to the problem.  Due to the fact that the gas
is polytropic, this relation is the required solution to the problem.

   In order to get a solution more adequate to the general case, we
can approximate the whole solution under the assumption that $V_c$
is a quantity of the first order, so:

\begin{gather}
    p_3 = p_{3_0} + p_3^\star,    \label{eq.8.a} \\
    V_3 = V_{3_0} + V_3^\star,    \label{eq.8.b} \\
    V_{3'} = V_{3'}^\star,       \label{eq.8.c}
\end{gather}

\noindent where the quantities with a star are of the first
order and the subscript $0$ represents the values at zeroth order
approximation.  Substitution of eqs.(\ref{eq.8.a})-(\ref{eq.8.c}) into
eqs.(\ref{eq.6.b})-(\ref{eq.6}) gives:

\begin{gather}
   v_2^2 = v_{2o}^2 - V_3^\star (p_{3_0} - p_2) + p_3^\star
             ( V_2 -V_{3_0} ),                 \label{eq.8.d} \\
   v_c^2 = (p_{3_0} - p_1)(V_c-V_{3'}^\star).  \label{eq.8.e}
\end{gather}

\noindent From the shock adiabatic relation, eq.(\ref{eq.2}), and
eqs.(\ref{eq.8.a})-(\ref{eq.8.c}) it follows that

\begin{gather}
   { \frac{ V_{3_0} }{V_2} } = { \frac{ (\gamma + 1)p_2 +
            (\gamma-1)p_{3_0} }{ (\gamma - 1)p_2 + (\gamma +
            1)p_{3_0} } },              \label{eq.8.f} \\
   { \frac{ V_{3'}^\star }{V_c} } = { \frac{(\gamma_c + 1)p_1 +
            (\gamma_c - 1)p_{3_0} } { (\gamma_c-1)p_1 +
            (\gamma_c+1)p_{3_0} } },    \label{eq.8.g} \\
   { \frac{ V_3^\star}{V_2} } = - { \frac{4\gamma p_2 p_3^\star}{
              \left[ (\gamma-1)p_2 + (\gamma+1)p_{3_0} \right]^2
            } }.                       \label{eq.8.h} 
\end{gather}

\noindent Substitution of eqs.(\ref{eq.8.d})-(\ref{eq.8.e}) and
eq.(\ref{eq.8.h}) in eq.(\ref{eq.5}) gives the required solution:

\begin{gather}
   { \frac{p_3^\star}{p_2} }= -{  \frac{V_c}{V_2}  \left(
               \frac{ \vert{\alpha}\vert + \beta }{\eta} \right) },
							\label{eq.9} \\
\intertext{where:}
   \beta = \left( {\frac{ p_{3_0} }{p_2} } - { \frac{p_1}{p_2} } \right)
            \left( 1 - \frac{ V_{3'}^\star }{V_c}  \right),  
            						\notag \\
   \eta = { \left(1-{ \frac{ V_{3_0} }{V_2} } \right) - \left( {
            \frac{ p_{3_0} }{p_2} } -1 \right) { \frac{ (\gamma -1) - (\gamma +
            1)V_{3_0}/V_2 }{ (\gamma -1) + (\gamma +
            1)p_{3_0}/p_2 } } },  
            						\notag \\
   \vert{\alpha}\vert^2 = 4 \frac{V_2}{V_c} \left(
            \frac{ p_{3_0} }{p_2} - 1 \right)\! \left( \frac{ p_{3_0} }{p_2} -
            \frac{p_1}{p_2} \right)\! \left( 1- \frac{ V_{3_0} }{V_2} \right)
            \!\left( 1- \frac{ V_{3'}^\star }{V_c} \right). 
            						\notag
\end{gather}

\begin{figure}
   \begin{center}
       \includegraphics[height=8.4cm]{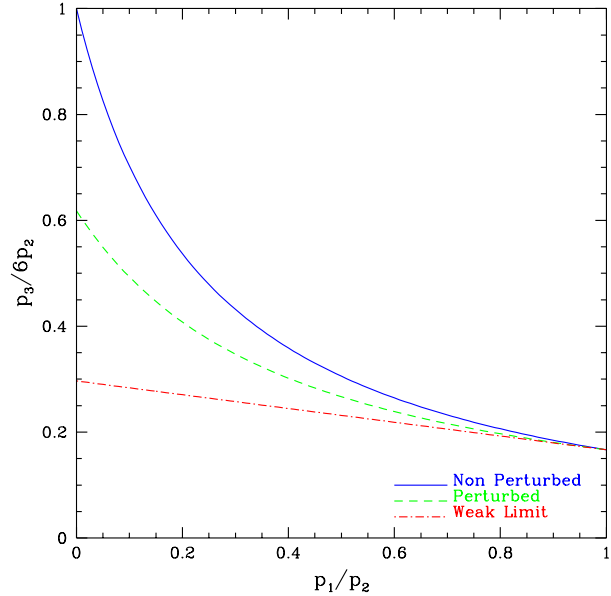}
   \end{center}
   \caption{ \footnotesize Variation of the pressure $p_3$ behind
	    the transmitted shock into the cloud as a function of the
	    strength of the initial incident shock.  The continuous
	    line shows the case for which the cloud is a solid wall with
	    infinite density.  The dashed curve is the solution at first
	    order approximation for which the cloud's specific volume is
	    a quantity of the first order.  The acoustic approximation
	    for which the incident shock is weak, at the same order of
	    accuracy, is represented by a dot-dashed curve. The perturbed
	    solutions were plotted assuming $\rho_c/\rho_1\!=\!100$ for
	    polytropic indexes $\gamma \!  = \! \gamma_c \! = \! 5/3$,
	    corresponding to a monoatomic gas.}
\label{fig.2}
\end{figure}

\noindent The specific volumes $V_{3_0}$ and $V_{3'}^\star$ are given
by eq.(\ref{eq.8.f}) and eq.(\ref{eq.8.g}) respectively.  For
completeness, approximations to eq.(\ref{eq.9}) for the case of a very
strong incident shock and that of a weak incident shock are given:

\begin{gather}
   \frac{p_3^\star}{p_2} = - { { \frac{ 4\gamma^2 (\gamma+1) }
	   { (3\gamma-1)(\gamma-1)^2 } \frac{V_c}{V_1} } \left(
           \frac{3\gamma -1}{\gamma_c+1} \, + \, \kappa \right), }
							\label{eq.10} \\
   \frac{ p_3^\star}{p_1} = - { 2 \zeta \frac{\gamma}{\gamma_c}
            \sqrt{ \frac{V_c}{V_1} } \left( \sqrt{
            \frac{\gamma_c}{\gamma} } + \sqrt { \frac{V_c}{V_1} }
            \right), }  
							\label{eq.11} \\
   \intertext{where:}
   \kappa = 2\, \sqrt{ \frac{V_1}{V_c} { \frac{ (3\gamma-1) (\gamma-1)
             }{ (\gamma_c+1) (\gamma+1) } } } \, ,
             						\notag 
\end{gather}

\noindent and $\zeta \! \equiv \! (p_2 \! - \!  p_1)/p_1 \! \ll 1 $ in
the weak limit.  Fig.(\ref{fig.2}) shows a plot of the pressure $p_3$ as a
function of the strength of the incident shock.  It is interesting to note
that even for very strong incident shocks the ratio $p_3/p_2$ differs from
zero, which follows directly from eq.(\ref{eq.7}) and eq.(\ref{eq.10}).
This simple means that the reflected shock is not strong, no matter the
initial conditions chosen.

%%%%%%%%%%%%%%%%%%%%%%%%%%%%%%%%%%%%%%%%%%%%%%%%%%%%%%%%%%%%%%%%%%%%
%%%%%%%%%%%%%%%%%%%%%%%%%%%F I G U R E 2%%%%%%%%%%%%%%%%%%%%%%%%%%%%
%%%%%%%%%%%%%%%%%%%%%%%%%%%%%%%%%%%%%%%%%%%%%%%%%%%%%%%%%%%%%%%%%%%%
%%%%%%%%%%%%%%%%%%%%%%%%%%%%%%%%%%%%%%%%%%%%%%%%%%%%%%%%%%%%%%%%%%%%
%%%%%%%%%%%%%%%%%%%%%%%%%%%F I G U R E 2%%%%%%%%%%%%%%%%%%%%%%%%%%%%
%%%%%%%%%%%%%%%%%%%%%%%%%%%%%%%%%%%%%%%%%%%%%%%%%%%%%%%%%%%%%%%%%%%%

   There are certain important general relations for which the above
results are a consequence of.	Firstly, by definition the pressure $p_2$
behind the shock is greater than the pressure $p_1$ of the environment.
Now consider a strong incident shock, then since $p_3 \! > \! p_2
\! \gg \!  p_1$, it follows that the transmitted shock into the cloud
is very strong.  Also, the reflected shock does not have to compress
too much the gas behind it to acquire the required equilibrium, so
it is not a strong shock. This last statement is in agreement with
eq.(\ref{eq.10}).  In general, for any strength of the incident shock,
since the inequality $p_3 \! > \! p_2 \! > \!  p_1$ holds, continuity
demands that the reflected shock can not be strong and, more importantly,
that the penetrating shock is always stronger than the reflected one.

   Secondly, very general inequalities are satisfied by the velocities
$v_2$, $v_c$, $v_{sl}$ as defined in fig.(\ref{fig.1}).  For instance:

\begin{equation}
   v_{sl} \! > \! v_2.
\label{eq.12}
\end{equation}

\noindent Indeed, since $V_3 v_2 / (V_3 \! - \!  V_2) \!  > \! v_2$ holds,
and the left hand side of this inequality is just $v_{sl}$ according to
mass flux conservation across the reflected shock, the result follows.

   On the other hand, from eq.(\ref{eq.6.b}) and eq.(\ref{eq.6}), since
$p_2 \! > \! p_1$ it follows that a necessary and sufficient condition for
$v_2 \! > \!  v_c$ to be true is that $V_2 \! - \!  V_3 \! > \! V_c \!
- \!  V_{3'}$.  This last condition is satisfied for sufficiently small
values of $V_c$.  To give an estimate of the smallness of the cloud's
specific volume needed, note that a necessary and sufficient condition
for $V_2 \! - \!  V_3 \! > \! V_c \!  - \!  V_{3'}$ to be valid is:

\begin{equation}
   { \frac{ V_2 (p_3 - p_2) }{ (\gamma-1)p_2 + (\gamma+1)p_3 } } > {
           \frac{ V_c (p_3-p_1) }{ (\gamma_c-1)p_1 + (\gamma_c+1)p_3 } },
\label{eq.12.a}
\end{equation}          

\noindent according to the shock adiabatic relation for the transmitted
and reflected shocks.  Since $p_3 \! > \! p_2 \! > \! p_1$ and $V_2 \! < \!
V_1$ it follows that: 

\begin{equation}
   \frac{V_1}{ (\gamma-1) + (\gamma+1)p_3/p_1 }  > \frac{V_c} {
               (\gamma_c - 1) + (\gamma_c+1)p_3/p_1} , 
\label{eq.12.b}
\end{equation}

\noindent which is very similar to eq.(\ref{eq.1}). In the same fashion,
under the assumption that the polytropic indexes are of the same order of
magnitude, eq.(\ref{eq.12.b}) implies $V_c\!<\!V_1$, which was an initial
assumption. Although eq.(\ref{eq.12.b}) is not sufficient, due to the
fact that $V_c$ is a first order quantity we can use in what follows:

\begin{equation}
   v_2 > v_c.
\label{eq.13}
\end{equation}

   The inequalities in eq.(\ref{eq.12}) and eq.(\ref{eq.13}) will prove
to be useful later when we choose a more suitable reference system to
describe the problem in question.

\section{Second initial discontinuity}

   Let us now  analyse the situation for which $0 \! < \! t\! < \! \tau$.
To begin with let us prove that:

\begin{equation}
   w_1 < v_2 + v_c \equiv u_2 ,
\label{eq.14}
\end{equation}

\noindent where the velocities $w_1$, $v_2$ and $v_c$ are defined in
fig.(\ref{fig.1}).  Suppose that the inequality in eq.(\ref{eq.14})
is not valid, then, by expressing the velocities as function of the
specific volumes and pressures by means of eq.(\ref{eq.3}) and the fact
that $p_2 \! > \! p_1$, $p_3 \! > \! p_4$ and $V_{4'} \! > \! V_3$,
it follows that $\rho_3 \! > \!  \rho_c$; then as the cloud's density
grows without limit, so does $\rho_3$.  Necessarily, eq.(\ref{eq.14})
has to be valid for sufficiently small values of the cloud's specific
volume. It is important to point out that since $w_2 \! = \!  \vert u_2
\! - \!  w_1 \vert \!  = \! u_2 \! - \!  w_1 $, the gas in region 2
as drawn in fig.(\ref{fig.1}) travels in the positive $x$ direction.
According to fig.(\ref{fig.1}) flows in region $1$ and $3$ are related by

\begin{equation}
   w_1 - w_3 = v_c,
\label{eq.15}
\end{equation}

   	Let us now prove a very general property of the
solution. Regions $2$ and $3$ are related to one another by the shock
adiabatic relation. Since the gas in regions $3'$ and $4$ obey a
polytropic equation of state $p_3 / p_4 \! = \! \left( V_4 / V_{3'}
\right)^{\gamma_c}$, it follows that:

\[
   \frac{p_4}{p_2} = { { \left( \frac{ V_{3'} }{V_4}\right)^{\gamma_c}
          } { \frac{ (\gamma+1)V_2 - (\gamma-1)V_3 }{ (\gamma+1)V_3 -
          (\gamma-1)V_2 } } }.
\]

\noindent Now, due to the fact that $V_{3'} \! < \! V_4 \! < \! V_1 $
$V_3 \! < \! V_2 \! < \! V_1 $ and $\gamma, \gamma_c \! > \!  1$ for
a reasonable equation of state, this relation can be brought to the form

\begin{equation}
   \frac{p_4}{p_2} < {  \frac{1}{2} \left[-(\gamma-1)+(\gamma+1)
		     \frac{V_2}{V_1} \right]  \to  0 \, , \ \  {\rm as} \;
                    \frac{p_1}{p_2} \to 0 }
\label{eq.15b}
\end{equation} 

\noindent according to the shock adiabatic relation.  This result implies
that most of the energy from the incoming shock has been injected to
the cloud, no matter how strong the initial incident shock is. Only a
very small amount of this energy is transmitted to the external gas that
lies in the other side of the cloud.  Note that this result is of a very
general nature since no assumptions about the initial density contrast
of the environment were made.

   In order to continue with a solution at first order approximation
in $V_c$, note that we have to use eqs.(\ref{eq.8.a})-(\ref{eq.8.c})
together with:

\begin{gather}
   p_4 = p_1 + p_4^\star ,       \label{eq.15.a} \\
   V_4 = V_4^\star ,             \label{eq.15.b} \\
   V_{4'} = V_1 + V_{4'}^\star,  \label{eq.16}
\end{gather}

\noindent where the quantities with a star are of the first order.
The velocities $w_1$ and $w_3$  can be expressed as functions of
the specific volumes and pressures by means of eq.(\ref{eq.3}), for
which after substitution of eqs.(\ref{eq.15.a})-(\ref{eq.16}) it
follows:

\begin{gather}
   w_1^2 = -p_4^\star V_{4'}^\star ,  \label{eq.16.a} \\
   w_3 = { { \frac{2}{\gamma_c - 1} } \left( \sqrt{\gamma_c
            p_{3_0} V_{3'}^\star } - \sqrt{\gamma_c p_1 V_4^\star}
            \right) }.    	      \label{eq.16.b}
\end{gather}                        
  
   The specific volumes behind the transmitted shock and the
reflected rarefaction wave are obtained from the shock adiabatic
relation and the polytropic equation of state for the gas inside the
rarefaction wave:

\begin{gather}
   V_{4'}^\star = -V_1 \frac{ p_4^\star }{ \gamma p_1 } ,
           		              \label{eq.16.c} \\
   V_4^\star = V_{3'}^\star \left( \frac{ p_{3_0} }{ p_{1} }
           \right)^{1/\gamma_c}.        \label{eq.16.d}
\end{gather}  

\noindent By substitution of eqs.(\ref{eq.16.a})-(\ref{eq.16.d}) and
eq.(\ref{eq.6}) in eq.(\ref{eq.15}) the required solution is found:

\begin{gather}
    \frac{p_4^\star}{p_2} = { \sqrt{ \frac{\gamma p_1}{p_2}
        \frac{V_c}{V_1} } \, \biggl( \Gamma + \Psi \Lambda \biggr), }
							\label{eq.17} \\
\intertext{where:}
   \Psi=\frac{ 2 \sqrt{\gamma_c} }{\gamma_c - 1} { \sqrt{ \frac
	{ (\gamma_c + 1)p_1/p_2 + (\gamma_c - 1)p_{3_0}/p_2 }
	{ (\gamma_c - 1)p_1/p_2 + (\gamma_c + 1)p_{3_0}/p_2 } } } \, , 
							\notag \\
   \Gamma=\frac{ \sqrt{2}\, (p_{3_0} - p_1)/p_2 }{ \sqrt
        { (\gamma_c-1)p_1/p_2 + (\gamma_c+1)p_{3_0}/p_2 } } \, ,        
        						\notag \\
   \Lambda=\sqrt{ \frac{ p_{3_0} }{p_2} } - \sqrt{ \frac{p_1}{p_2}
         \left( \frac{p_{3_0}}{p_{1}} \right)^{1/\gamma_c} }\, .
         						\notag
\end{gather}

   For completeness the limits for the case of strong and weak incident
shocks are given:

\begin{gather}
   \frac{p_4^\star}{p_2} = { \sqrt{ \frac{ \gamma (3\gamma-1) }
       { (\gamma_c+1) (\gamma-1) } \frac{p_1}{p_2} \frac{V_c}{V_1} }\,
       \Biggl( \sqrt{2} + \xi \Biggr) },
							\label{eq.18} \\
   \frac{p_4^\star}{p_2} = { 6 \zeta \sqrt{ \frac{\gamma}{
            \gamma_c } }\sqrt{ \frac{V_c}{V_1} } },  
							\label{eq.19} \\
\intertext{with:}
   \xi=\frac{ 2 \sqrt{\gamma_c} }{ \sqrt{ (\gamma_c - 1) } } \left[ 1
	 - \left( \frac{p_1}{p_2} \frac{\gamma-1}{3\gamma-1} \right)^
	 { (\gamma_c-1) / { 2 \gamma_c } } \right].
	 						\notag
\end{gather}

\begin{figure}
   \begin{center}
       \includegraphics[height=8.4cm]{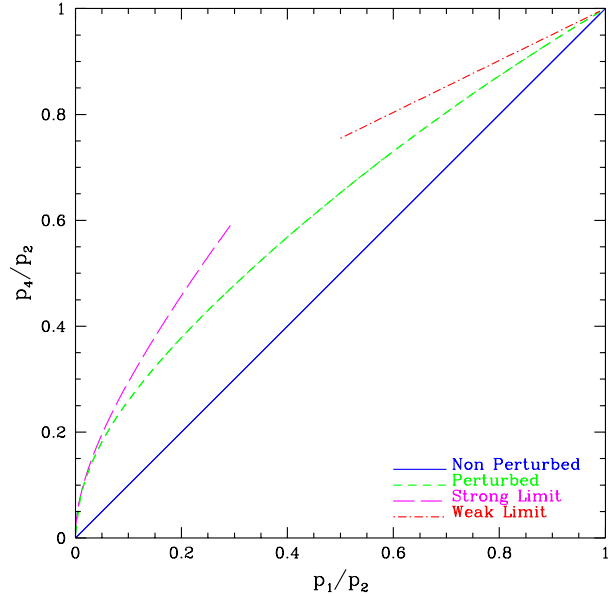}
   \end{center}
    \caption{ \footnotesize  Variation of the pressure $p_4$ behind the
             transmitted shock into the external medium as a function
             of the strength of the incident shock.  The continuous
             line represents the case for which the cloud has infinite
             density and so it does not transmit any shock to the external
             medium.  The dashed curve represents the case for which the
             cloud's specific volume is a quantity of the first order.
             The dash-dotted curve is the limit for which a strong
             (or weak) incident shock collides with the cloud at the
             same order of approximation.  The perturbed curves were
             produced under the assumption that $\rho_c/\rho_1\!=\!100$
             for monoatomic gases.}
\label{fig.3}
\end{figure}

     It follows from eq.(\ref{eq.18}) that $p_4 \! \ll \! p_2$ as the
strength of the incident shock increases without limit.  This result was
given by a very general argument in eq.(\ref{eq.15b}).  Fig.(\ref{fig.3})
shows the variation of the pressure $p_4$ behind the shock transmitted
to the environment as a function of the strength of the initial incident
shock, after the second initial discontinuity.

%%%%%%%%%%%%%%%%%%%%%%%%%%%%%%%%%%%%%%%%%%%%%%%%%%%%%%%%%%%%%%%%%%%%
%%%%%%%%%%%%%%%%%%%%%%%%%%%F I G U R E 3%%%%%%%%%%%%%%%%%%%%%%%%%%%%
%%%%%%%%%%%%%%%%%%%%%%%%%%%%%%%%%%%%%%%%%%%%%%%%%%%%%%%%%%%%%%%%%%%%
%%%%%%%%%%%%%%%%%%%%%%%%%%%%%%%%%%%%%%%%%%%%%%%%%%%%%%%%%%%%%%%%%%%%
%%%%%%%%%%%%%%%%%%%%%%%%%%%F I G U R E 3%%%%%%%%%%%%%%%%%%%%%%%%%%%%
%%%%%%%%%%%%%%%%%%%%%%%%%%%%%%%%%%%%%%%%%%%%%%%%%%%%%%%%%%%%%%%%%%%%

\section{General solution}

   Having found values for the pressures $p_3^\star$ and $p_4^\star$
as a function of the initial conditions $p_1$, $p_2$, $V_1$ and $V_c$,
the problem is completely solved.  Indeed, using the shock adiabatic
relation $V_2$ is known. With this, the values of $V_{3'}^\star$,
$V_{3}^\star$, $V_4^\star$ and $V_{4'}^\star$ are determined by
means of eq.(\ref{eq.8.g}), eq.(\ref{eq.8.h}), eq.(\ref{eq.16.c})
and eq.(\ref{eq.16.d}) respectively.  The complete values for
pressure and specific volumes are obtained thus with the aid of
eqs.(\ref{eq.8.a})-(\ref{eq.8.c}) and eqs.(\ref{eq.15.a})-(\ref{eq.16}).
The velocities of the flow, as defined in fig.(\ref{fig.1}), are
calculated either by mass flux conservation on crossing a shock, or by
the formula given for the velocity discontinuity in eq.({\ref{eq.3}).
The hydrodynamical values of the pressure $p_R$ and density $\rho_R$
inside the rarefaction wave come from eqs.(\ref{eq.4.a})-(\ref{eq.4}).

   In order to analyse the variations of the hydrodynamical quantities
as a function of position and time, let us now describe the problem
in a system of reference in which the gas far away to the right
of the cloud is always at rest, as presented in fig.(\ref{fig.4}).
Let $x_{tl}$ and $x_{tr}$ be the coordinates of the left and right
tangential discontinuities, $x_{sl}$ and $x_{sr}$ the coordinates of
the reflected and transmitted shocks produced after the first initial
discontinuity, $\chi_{sr}$ the position of the transmitted shock after
the second initial discontinuity and $x_a$ and $x_b$ the left and right
weak discontinuities which bound the rarefaction wave. The new velocities
are defined by Galilean transformations:
\begin{gather}
   u_2 = v_2 + v_c,           \label{eq.20} \\
   u_{sl} = v_{sl} - v_c,     \label{eq.21} \\
   u_{sr} = v_{sr} + v_c,     \label{eq.22} \\
   \nu_R = w_1 - w_R,         \label{eq.23} \\
   \nu_{sr} = w_1 + w_{sr}.   \label{eq.24} 
\end{gather}

    The direction of motion of the flow is shown in fig.(\ref{fig.4}) and
it follows from eq.(\ref{eq.12}), eq.(\ref{eq.13}) and eq.(\ref{eq.21})
that $u_{sl}$ points to the left in this system of reference.  Since,
in the same frame, $v_c$ and $w_1$ point to the right, continuity
across a weak discontinuity demands $\nu_R$ to do it in the same way.

%%%%%%%%%%%%%%%%%%%%%%%%%%%%%%%%%%%%%%%%%%%%%%%%%%%%%%%%%%%%%%%%%%%%
%%%%%%%%%%%%%%%%%%%%%%%%%%%F I G U R E 4%%%%%%%%%%%%%%%%%%%%%%%%%%%%
%%%%%%%%%%%%%%%%%%%%%%%%%%%%%%%%%%%%%%%%%%%%%%%%%%%%%%%%%%%%%%%%%%%%
\begin{figure}
   \begin{center}
       \includegraphics[height=5.4cm]{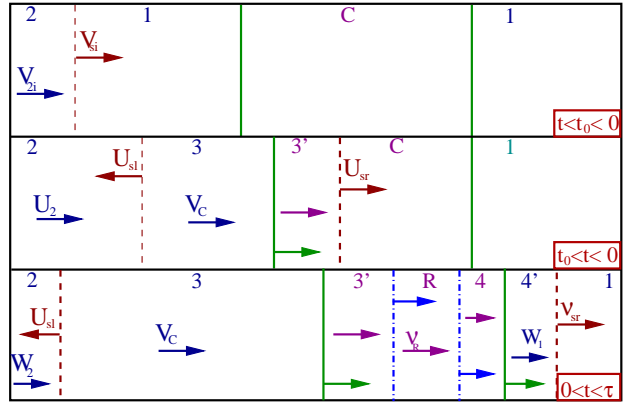}
   \end{center}
    \caption{ \footnotesize Description of the problem of a collision
             of a shock with a cloud in a system of reference for which
             the gas far away to the right (at $x \! = \! \infty$) is
             always at rest. Originally a shock is travelling to the
             right and hits a tangential discontinuity (top panel).
             This produces a discontinuity in the initial conditions
             so a reflected and transmitted shock are produced;  the
             gas in the cloud begins to accelerate (middle panel).
             Eventually the transmitted shock into the cloud collides with
             its right boundary producing a ``reflected'' rarefaction
             wave bounded by two weak discontinuities (region R) and a
             transmitted shock into the external medium (lower panel).
             In this system of reference every single discontinuity
             produced by means of the interaction move to the right,
             except for the reflected shock produced after the first
             collision.  Arrows represent the direction of motion of
             various boundaries and direction of flow.  Numbers label
             different regions of the flow.  Dashed lines represent shock
             waves, dash-dotted ones weak discontinuities and continuous
             ones tangential discontinuities.}
\label{fig.4}
\end{figure}
%%%%%%%%%%%%%%%%%%%%%%%%%%%%%%%%%%%%%%%%%%%%%%%%%%%%%%%%%%%%%%%%%%%%
%%%%%%%%%%%%%%%%%%%%%%%%%%%F I G U R E 4%%%%%%%%%%%%%%%%%%%%%%%%%%%%
%%%%%%%%%%%%%%%%%%%%%%%%%%%%%%%%%%%%%%%%%%%%%%%%%%%%%%%%%%%%%%%%%%%%

   The tangential discontinuities and the shocks produced by the
initial discontinuities move with constant velocity throughout the gas.
This implies that the time at which the first initial discontinuity
takes place is:

\begin{equation}
   t_0 = - { \frac{\Delta}{ u_{sr} } }, 
\label{eq.25}
\end{equation}

\noindent where $\Delta$ represents the initial width of the cloud. Hence,
the positions of all different discontinuities for $t_0 \! < \!  t \! <
\! 0$ are:

\begin{gather}
   x_{sr} = u_{sr}t,               \label{eq.26} \\
   x_{sl} = { -\Delta - u_{sl} (t - t_0) },  \label{eq.27} \\
   x_{tl} = { -\Delta + v_c  ( t  - t_0) }.  \label{eq.28}
\end{gather}

\noindent and for $0 \! < \! t \! < \! \tau$,
eqs.(\ref{eq.27})-(\ref{eq.28}) are valid together with

\begin{gather}
   x_{a} = { -t \left(  \frac{\gamma_c+1}{2}  w_3 + c_4
            \right) + w_1 t },              \label{eq.29} \\
   x_b = \left( w_1 - c_4 \right) t,        \label{eq.30} \\
   \chi_{sr} = \nu_{sr} t,                  \label{eq.31} \\
   x_{tr} = w_1 t.                          \label{eq.32}
\end{gather}

\noindent The time $\tau$ at which the left tangential discontinuity
collides with the left boundary of the rarefaction wave is given by
$x_{tl} \! = \! x_a$, and thus:

\begin{equation}
   \tau c_{3'} = { {v_c t_0} + \Delta}.
\label{eq.33}
\end{equation}

\noindent Fig.(\ref{fig.5}) shows the variation of the pressure and
density as a function of time and position in a system of reference in
which the gas far away to the right of the cloud is at rest.

%%%%%%%%%%%%%%%%%%%%%%%%%%%%%%%%%%%%%%%%%%%%%%%%%%%%%%%%%%%%%%%%%%%%
%%%%%%%%%%%%%%%%%%%%%%%%%%%F I G U R E 5%%%%%%%%%%%%%%%%%%%%%%%%%%%%
%%%%%%%%%%%%%%%%%%%%%%%%%%%%%%%%%%%%%%%%%%%%%%%%%%%%%%%%%%%%%%%%%%%%
\begin{figure*}
      \begin{center}
       \includegraphics[width=9cm]{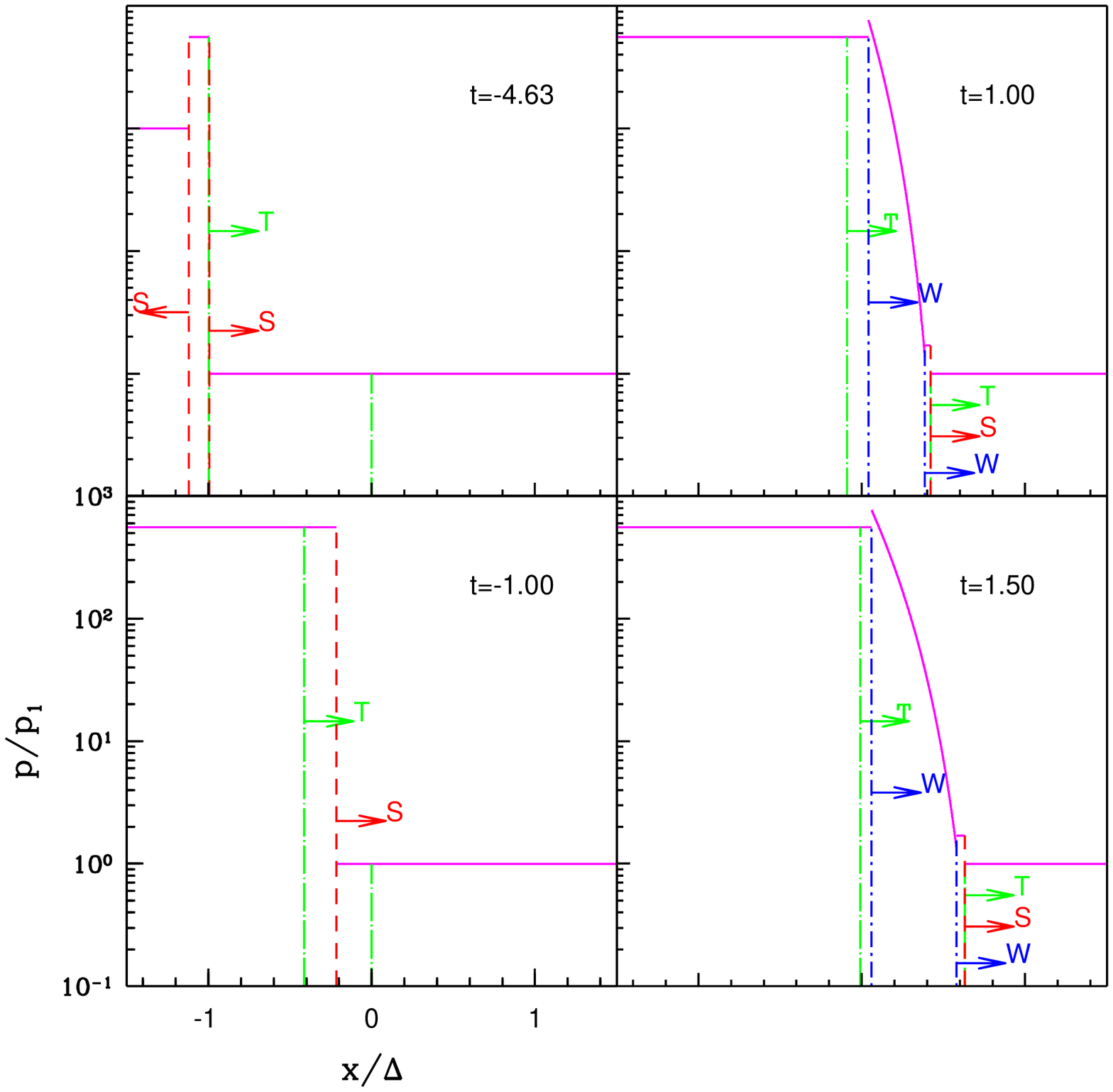}
       \includegraphics[width=9cm]{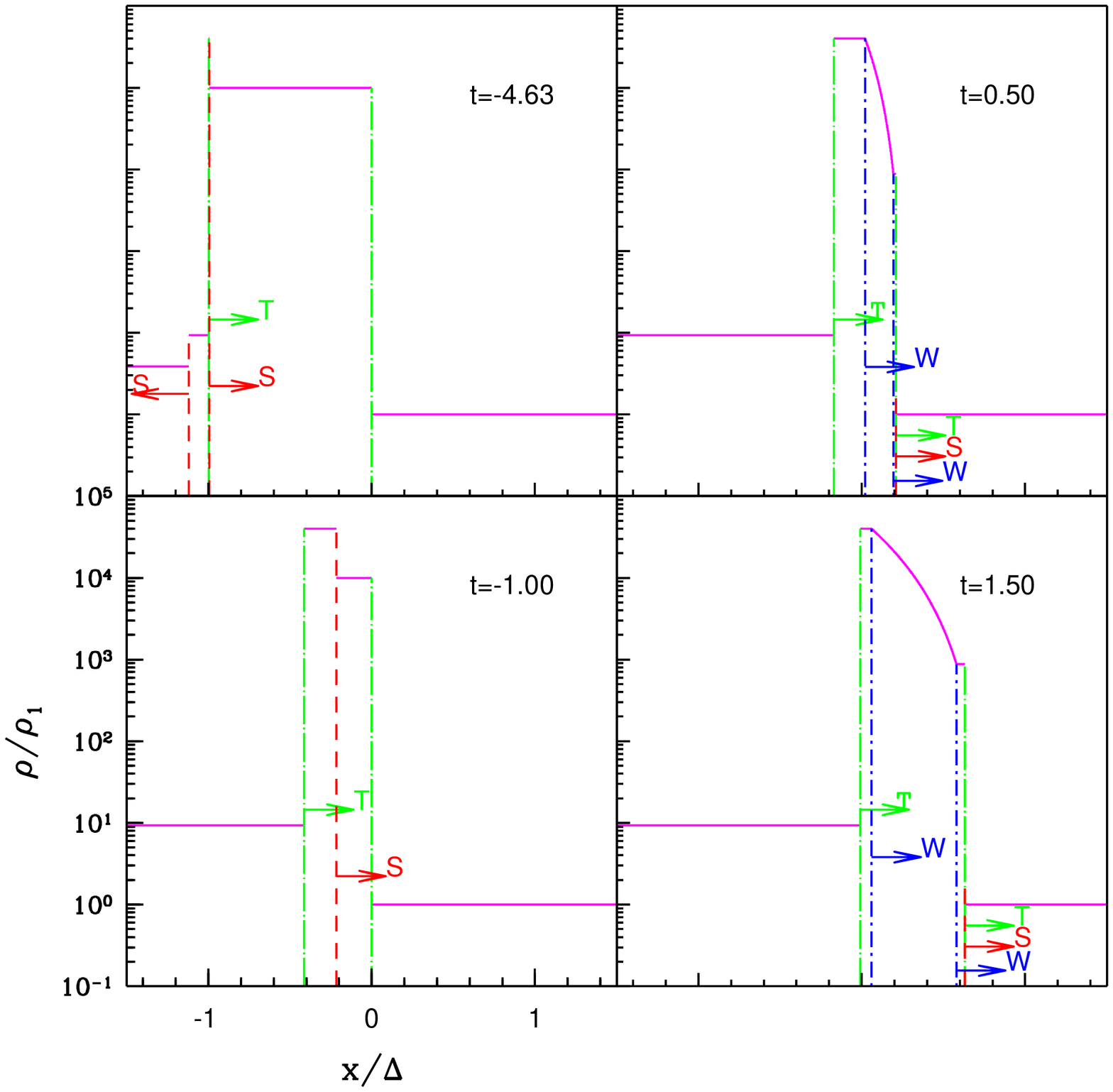}
   \end{center}
    \caption{ \footnotesize Variation of the pressure $p$ and density
             $\rho$ (with respect to the initial pressure $p_1$
             and density $\rho_1$ of the environment) as a function
             of position $x$ (normalised to the initial width of the
             cloud $\Delta$) and dimensionless time $t$ (in units of the
             time $\Delta/c_1$ --where $c_1$ is the speed of sound in
             the external medium).  Dashed lines represent shock waves
             (S), dot-dashed lines are tangential discontinuities (T),
             which are boundaries of the cloud,  and short-long dashed
             lines represent weak discontinuities (W), which bound a
             rarefaction wave.  The system of reference was chosen so that
             gas far away to the right of the diagram is at rest. The
             diagram shows the case for which $\rho_c/\rho_1\!=\!10^4$,
             and the polytropic indices correspond to a monoatomic gas.}
\label{fig.5}
\end{figure*}
%%%%%%%%%%%%%%%%%%%%%%%%%%%%%%%%%%%%%%%%%%%%%%%%%%%%%%%%%%%%%%%%%%%%
%%%%%%%%%%%%%%%%%%%%%%%%%%%F I G U R E 5%%%%%%%%%%%%%%%%%%%%%%%%%%%%
%%%%%%%%%%%%%%%%%%%%%%%%%%%%%%%%%%%%%%%%%%%%%%%%%%%%%%%%%%%%%%%%%%%%

   The width of the cloud varies with time, and it follows from
eq.(\ref{eq.28}) and eq.(\ref{eq.32}) that this variation is given by:

\begin{equation}
   \bar{X}(t) = \Theta(t) w_1 t + \Delta - v_c (t-t_0),
\label{eq.34}
\end{equation}

\noindent where $\Theta(t)$ is the Heaviside step function. This
linear relation is plotted in fig.(\ref{fig.7}).

%%%%%%%%%%%%%%%%%%%%%%%%%%%%%%%%%%%%%%%%%%%%%%%%%%%%%%%%%%%%%%%%%%%%
%%%%%%%%%%%%%%%%%%%%%%%%%%%F I G U R E 6%%%%%%%%%%%%%%%%%%%%%%%%%%%%
%%%%%%%%%%%%%%%%%%%%%%%%%%%%%%%%%%%%%%%%%%%%%%%%%%%%%%%%%%%%%%%%%%%%
\begin{figure}
   \begin{center}
       \includegraphics[height=8.4cm]{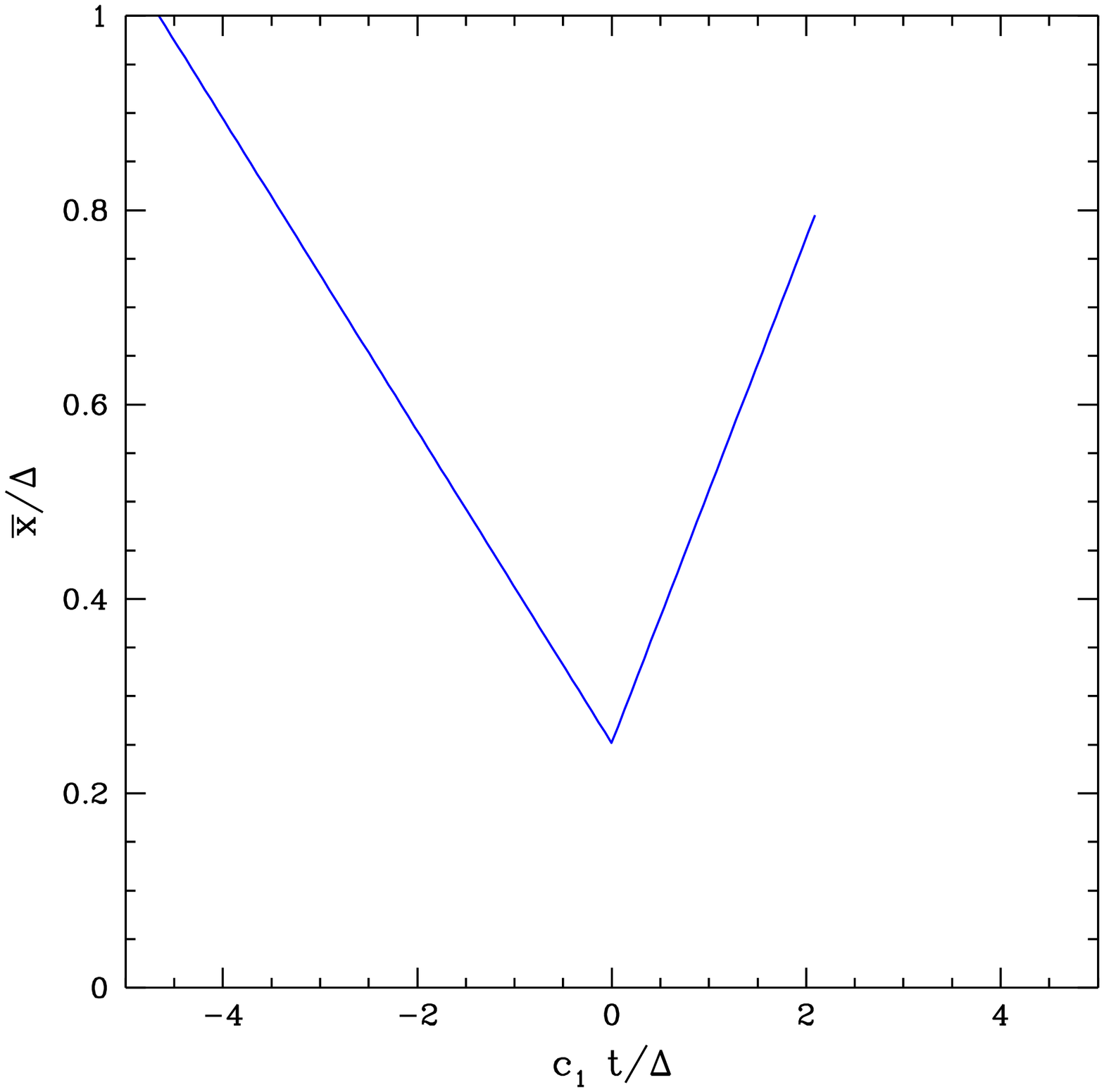}
   \end{center} 
    \caption{ \footnotesize Variation of the width of the cloud in
             units of its original size $\Delta$ as a function of the
             dimensionless quantity $c_1t/\Delta$.  Where $c_1$ represents
             the sound speed of the gas for the external environment and
             $t$ the time. The curve was produced under the assumption
             that $\rho_c/\rho_1\!=\!10^4$.  The gas was considered to
             be monoatomic.}
\label{fig.7}
\end{figure}
%%%%%%%%%%%%%%%%%%%%%%%%%%%%%%%%%%%%%%%%%%%%%%%%%%%%%%%%%%%%%%%%%%%%
%%%%%%%%%%%%%%%%%%%%%%%%%%%F I G U R E 6%%%%%%%%%%%%%%%%%%%%%%%%%%%%
%%%%%%%%%%%%%%%%%%%%%%%%%%%%%%%%%%%%%%%%%%%%%%%%%%%%%%%%%%%%%%%%%%%%

\section{Summary}

   The problem of a collision of a plane parallel shock wave with a high
density cloud bounded by two plane parallel tangential discontinuities has
been discussed.  Radiation losses, magnetic fields and self gravity of
the cloud were neglected. General analytic solutions were found for the
simple case in which the ratio of the environment's density to that of
the cloud's density is a quantity of the first order.

  When the shock collides with the boundary of the cloud, a discontinuity
in the initial conditions is produced.  This splits the incoming shock
into two shock waves: one which penetrates the cloud and one which is
reflected back to the external medium.  When the transmitted shock into
the cloud reaches the opposite boundary, another discontinuity in the
initial conditions is produced, causing the transmission of a shock
wave to the external medium and the reflection of a rarefaction wave
from the point of collision.

\section{Acknowledgements}

   I would like to thank Malcolm Longair for fruitful comments and Paul
Alexander for useful discussions while doing this work.  I thank support
granted by Direcci\'on General de Asuntos del Personal Acad\'emico
(DGAPA) at the Universidad Nacional Aut\'onoma de M\'exico (UNAM).

\bibliographystyle{abbrvnat}
\bibliography{cloud}

% \eject

% \pagestyle{empty}

% \null \vfill\eject

% \null \vfill\eject

% \null \vfill\eject

% \null \vfill\eject

\end{document}